\documentclass[runningheads]{llncs}

\usepackage{graphicx}
\usepackage{algorithm}
\usepackage[noend]{algpseudocode}

\newcommand{\startlist}{
\setlength{\itemsep}{0mm}
\setlength{\topsep}{0mm}
\setlength{\leftmargin}{6mm}
\setlength{\rightmargin}{0mm}
\setlength{\parskip}{0pt} \setlength{\parsep}{0pt}
}
\newcommand{\listlabel}{$\bullet$}

\newenvironment{myitemize}{
\begin{list}{\listlabel}{\startlist}
}{
\end{list}
}

\begin{document}

\title{System-aware dynamic partitioning for batch and streaming workloads\thanks{
This work was supported by project "Integrated program for training new generation of researchers in the disciplinary fields of computer science" (EFOP-3.6.3-VEKOP-16-2017-00002), by the European Union and co-funded by the European Social Fund.
}}

\author{\centerline{Zolt\'an Zvara\inst{1}
\and
P\'eter G.N. Szab\'o\inst{3}
\and
Bal\'azs Barnab\'as L\'or\'ant\inst{2}
\and
Andr\'as A. Bencz\'ur\inst{1}
}}

\authorrunning{Zvara et al.}

\institute{\centerline{Institute For Computer Science and Control,
E\"otv\"os L\'oránd Research Network (ELKH)}
\email{\{zoltan.zvara,\,benczur\}@sztaki.hu} \and
E\"otv\"os University Budapest, Doctoral School of Informatics
\email{babuafi@inf.elte.hu} \and
University of Pécs, Faculty of Engineering and Information Technology
\email{szabo.peter@mik.pte.hu}
}

\maketitle

\begin{abstract}
When processing data streams with highly skewed and nonstationary key distributions, we often observe overloaded partitions when the hash partitioning fails to balance data correctly. To avoid slow tasks that delay the completion of the whole stage of computation, it is necessary to apply adaptive, on-the-fly partitioning that continuously recomputes an optimal partitioner, given the observed key distribution.
While such solutions exist for batch processing of static data sets and stateless stream processing, the task is difficult for long-running stateful streaming jobs where key distribution changes over time. Careful checkpointing and operator state migration is necessary to change the partitioning while the operation is running.

Our key result is a lightweight on-the-fly Dynamic Repartitioning (DR) module for distributed data processing systems (DDPS), including Apache Spark and Flink, which improves the performance with negligible overhead.
DR can adaptively repartition data during execution using our Key Isolator Partitioner (KIP). In our experiments with real workloads and power-law distributions, we reach a speedup of 1.5-6 for a variety of Spark and Flink jobs.

\end{abstract}

\section{Introduction}
In the era of Big Data, complex, parallel data analytic tools for both batch and stream processing are essential for the industry and research.
A few years ago, the term \textbf{fast data}~\cite{lam2012muppet} arose to capture the idea that \textbf{streams} of data are generated at very high rates and that these need to be analyzed quickly to arrive at actionable intelligence~\cite{bifet2011data}.

Fast data is not just about processing power, but also about \textbf{fast changes in semantics}.
Large databases available for mining currently have been gathered over months or years, and the underlying processes that generate them have changed over time, sometimes radically~\cite{hulten2001mining,zliobaite2012next}. Examples include webshop traffic at Black Friday, a music recommendation system after an album release, or a news portal during an election or a disaster.
In data analysis tasks, fundamental properties of the data can change quickly, which makes gradual manual model adjustment procedures inefficient and even infeasible~\cite{zliobaite2012next}.
Traditional batch distributed data processing systems assume finite, static, identically distributed datasets. By contrast, stateful stream processors need to adapt to distributions that evolve. Processing strongly depends on the order of examples generated from a continuous, non-stationary flow of data. Performance is hence affected by potential statistical changes in the data called \textbf{concept drift}~\cite{gama2013evaluating}.

In this paper, we address the challenge of skewed key distributions that severely degrade the automatic scaling and load balancing capabilities of the DDPS.
We concentrate on tasks with concept drift where the set of heavy keys change in long-running stateful jobs. In such cases, it may be necessary to rebalance the keys several times without stopping and restarting the job.
Our goal is to address repartitioning needs in all data stream processing, be it micro-batch, such as Spark~\cite{zaharia2010spark}, or real streaming, such as Flink~\cite{carbone2015lightweight}.

The main technical difficulty of skew mitigation is an appropriate design for efficient integration with the underlying system. We design Dynamic Repartitioning as a pluggable framework that can be integrated with most MapReduce-based data processing frameworks. 
In the simplest tasks, such as counting, we can apply Map-side combiners to reduce the load of heavy keys in the next stage. We concentrate on more complex, stateful tasks, such as \emph{join} and \emph{groupBy}, where we cannot combine operations inside the Mapper. Examples of such tasks with high sensitivity to skewed key distribution include building an inverted search index and maintaining order in a windowed state.

To our best knowledge, our DR system is the first to integrate into two completely different DDPS frameworks, Spark and Flink.  By adding only a few hundred lines of DDPS-specific code, we give general APIs to interface with \textit{key grouping}, the general mechanism to process individual partitions on parallel operator instances. We also implement an efficient \textsl{replay} and \textsl{state migration} strategy close to the DDPS to handle keys assigned to a new reducer partition.

Our universal architecture fits both batch and stateful stream processing:
\begin{myitemize}
\item It adapts to concept drift in the distribution in a long-running data streaming process and introduces no additional latency, as there is no separate sampling job or pre-aggregation stage before it makes a partitioning decision.
\item It works for any operator, including complex stateful tasks, by state migration that existing streaming skew mitigation methods cannot handle.
\item It reuses normal DDPS communication, thus incurs minimal overhead.
\end{myitemize}

In our experiments, we obtained significant speedup in Spark and Flink tasks. For example, for random data with Zipf distribution, we reached 1.5--2.0 times speedup for both data streaming systems and 6 times on a web-crawl dataset.

Our main results are the following:
\begin{myitemize}
\item We demonstrate that after our system measures a small data portion, we can deploy replay (batch) and state migration (stream) as universal low-overhead solutions for mitigating data skew in modern distributed data processing systems. We measure speedup in all batch, micro-batch, and streaming environments with stateful operators.
\item We devise a low-memory-footprint sampling mechanism that, in combination with our hybrid hash function, provides better load balance with lower communication cost than in prior works~\cite{gedik2014partitioning}.
\end{myitemize}

The paper is organized as follows. After related results, in Section~\ref{sect:arch} we show the general design concept of DR.
Our drift respecting balanced hashing procedure is in Section~\ref{sect:hashing};  our low-memory distributed approximate heavy hitter counting method is in the  full paper.
In Sections~\ref{sect:eval} and~\ref{sect:webcrawl}, we evaluate the performance of our system over Spark and Flink with batch and stream processing tasks.
Our source codes and extended paper with additional details on the algorithms, implementation, and measurements are available at \url{https://github.com/zzvara/dynamic-repartitioning-paper}.

\section{Related works}
\label{sect:related}

\textbf{Database partitioning.} Results on database partitioning or sharding~\cite{schneider1989performance,zhang2015memory} concentrate on selecting the attribute for partitioning in conjunction with the SQL executor optimizer and use static, predetermined sampling. By contrast, we address the lower level execution after the developer or the DDPS optimizer selects the partitioning attribute, and we focus on the partitioning function.

\textbf{Skew in MapReduce.} The first related solutions mitigate reducer skew by measuring the key distribution during the Map operation~\cite{vernica2012adaptive,gufler2012partition}. In general, these methods are not appropriate for long-running streaming tasks with concept drifts in key distribution, nor for stateful operators that require state migration after re\-par\-ti\-tion\-ing.

The drawback of straggler detection methods~\cite{kwon2012skewtune,kwon2013managing} is that the same keys may end up in different partitions, hence they are not applicable to stateful operators or cannot be supported without extra merging steps.

Another possibility is to obtain count estimates by a static, predetermined sample typically run as a separate job~\cite{gates2013apache,metwally2012v}. These methods are best suited for large jobs with static key distributions: even if the underlying system is for data stream processing, frequent sampling would be too costly.

The main drawback of all Hadoop-based repartitioning techniques is either high migration~\cite{vernica2012adaptive,kwon2012skewtune,kwon2013managing} or  high sample counting cost~\cite{gates2013apache}.
In our micro-batch or data stream skew mitigation task, both measurement and reallocation must incur at least an order of magnitude lower cost than the micro-batch or the stream intra-checkpoint job.

\textbf{Skew in data streaming systems.} Other than the very early first results \cite{xing2005dynamic,abadi2005design}, all recent results focus on Storm \cite{toshniwal2014storm} or S4 \cite{neumeyer2010s4}. We are unaware of similar results in the most popular systems, Spark \cite{zaharia2010spark} and Flink \cite{carbone2015lightweight}.

The critical step for optimal balance 
is fission \cite{hirzel2014catalog}, the method to partition the stream to multiple workers. Fission is cumbersome for stateful operators such as \textit{groupBy}. Most solutions are unsafe \cite{fang2016parallel,hidalgo2017self} or at least strongly restricted in this sense. Safe data parallelism is considered in \cite{schneider2015safe} without runtime data distribution considerations. State migration cost is not considered in~\cite{katsipoulakis2017holistic}, while previous work~\cite{gedik2014partitioning,fang2016parallel} details its necessity in maintaining low latency.

Several results consider data skew in Storm. Nasir et al.~\cite{nasir2016two,nasir2015partial} perform streaming load balancing at data sources for the first stage of computation where the user-defined function (UDF) is an associative monoid (i.e.,\ Map-side combine is feasible).
Other works \cite{rivetti2015proactive,rivetti2015efficiently} ignore migration cost and sampling overhead, which are dominant costs in a long-running streaming job with frequent unpredictable concept drifts.

The experiments in~\cite{ding2015optimal} show that Storm (that just recently added support for stateful operations~\cite{toshniwal2014storm}) execution becomes faster despite a relatively costly optimization algorithm and a state migration process that relies on Zookeeper for worker synchronization. However, Zookeeper based coordination is too costly for Spark and Flink streaming execution that allows frequent repartitioning.

Related work on Storm uses external databases to store state, thus limited to simple word-count cases. Nevertheless, we make the best effort to enumerate and, as baselines, partly reconstruct the ideas in publications for S4 and Storm repartitioning. Experimentally comparing these systems against Spark and Flink is beyond the scope of this paper.

Unlike our solution, which dynamically repartitions stateful operators on-the-fly without state migration, most of the above results either find a balanced partitioning at the beginning of the distributed job \cite{ding2015optimal,rivetti2015proactive,rivetti2015efficiently} or repartition by reallocating keys, and do not apply to stateful operators \cite{nasir2016two,nasir2015partial,fang2016parallel,hidalgo2017self}.

\begin{figure*}[t]
   \centering
   \includegraphics[width=0.95\textwidth]{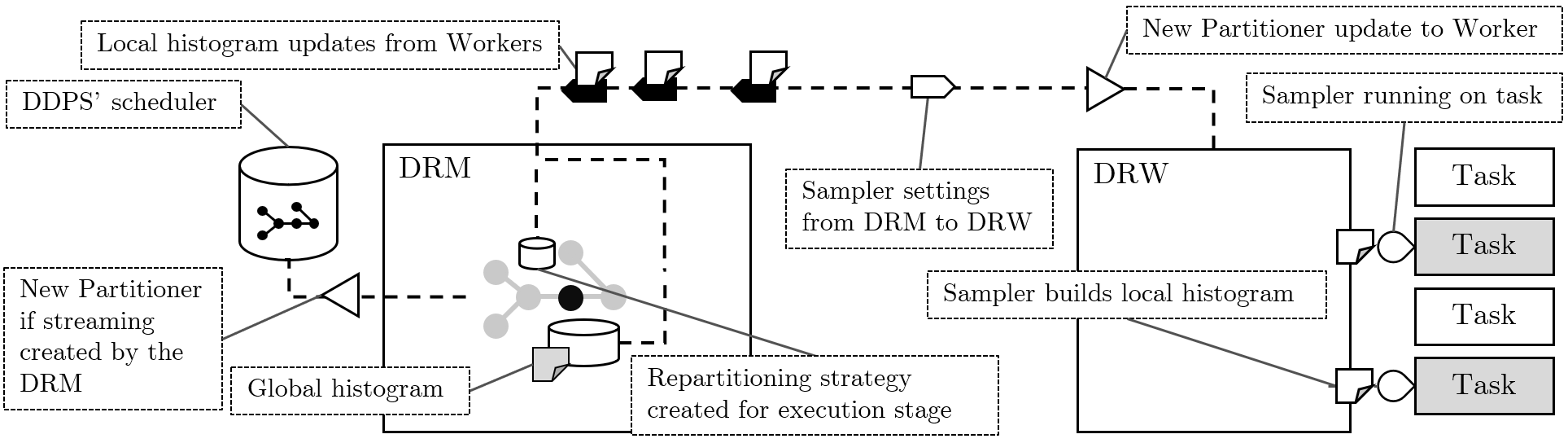}
   \caption{Integration of Dynamic Repartitioning with DDPS. Dashed lines represent communication between DR and DDPS components.}
   \label{fig:dr-arch}
\end{figure*}

\textbf{Partitioning and top-k approximation.} A first result in the area of identifying heavy hitters~\cite{dewitt1992practical} lists simple heuristics for partitioning and load scheduling. Algorithms such as low space heavy hitter identification~\cite{metwally2005efficient} were designed for data streams.

In our experiments, sketch algorithms and their variants are either only accurate for highly skewed data or consume unacceptable amounts of memory. We use Lossy Counting~\cite{manku2002approximate} and SpaceSaving~\cite{metwally2005efficient}, two key results in finding frequent items in data streams, as a baseline method in our experiments.

Gedik \cite{gedik2014partitioning} formalizes and develops partitioning functions for stateful operators based on a combination of consistent and explicit hashing. The author assumes a theoretical general-purpose stream processing system and considers the trade-off between migration cost and load balance. Our hash functions are closest to \cite{gedik2014partitioning}, but provide a better load balance as shown in our experiments.

\section{Architecture}
\label{sect:arch}

DR is a standalone library written in Scala, which is pluggable for any DDPS. Generic components are implemented in 2,000 lines of code, while the patches for Flink (an \emph{asynchronous} system) and Spark (a strictly \emph{synchronous} system) are under 500 lines each. Our architecture is shown in Figure~\ref{fig:dr-arch}. The Dynamic Repartitioning Master (DRM) is our central authority, integrated into the \textit{Driver} component of the DDPS. Dynamic Repartitioning Workers (DRW) are parts of the DDPS \textit{Worker}s.

There is a fundamental difference between the batch and the streaming operation in the timing and the cost of repartitioning. When we repartition a \textbf{batch} job, we may have to buffer the Mapper output after processing and use the new partitioning function as soon as it becomes ready. Ideally, we intervene while the data is still in the buffers and before it is evicted to the disk at the Mappers. Since during eviction, the system distributes data by using the actual hash partitioner, changing the partitioning function after data has been written to disk requires recomputing partition assignments (\textbf{replay}) using the new partitioner. Hence a batch job is repartitioned only in an early stage of the execution so that the cost of replay does not exceed the expected gains of better partitioning. Repartitioning actions in \textbf{streaming} can be taken at checkpoints or micro-batch boundaries. In stateful applications, repartitioning incurs \textbf{state migration}, hence the gains for repartitioning should exceed state migration costs. To ensure that a partitioner construction is useful in the long run, we keep a record of past histograms.

We implemented DR for \textbf{Spark} by injecting the partitioner into the job directed acyclic graph (DAG) that describes the streaming application. Due to the micro-batch nature of Spark Streaming, it uses the new partitioner when it generates micro-batches from the streaming DAG. Spark performs state migration automatically in the shuffle phase. In our \textbf{Flink} implementation, we make\hspace{0.15em}use\hspace{0.15em}of\hspace{0.15em}the\hspace{0.15em}Asynchronous\hspace{0.15em}Distributed\hspace{0.15em}Snapshot\hspace{0.15em}mech\-a\-nism~\cite{carbone2015lightweight} used for fault tolerance.

\section{The Key Isolator Partitioner}
\label{sect:hashing}

\def\KI{\mbox{KI}}
\def\Hash{\mbox{\textsc{Hash}}}
\def\Hist{\mbox{\textsc{Hist}}}
\def\maxload{\mbox{\textsc{maxload}}}
\def\hostload{\mbox{\textsc{hostload}}}

\begin{algorithm}[t]\small
\caption{\textsc{KIPUpdate} ($KI, \Hash, H, \Hist, N, \varepsilon$)}\label{algorithm:kip}
\begin{algorithmic}[1]
  \State $\maxload \gets \max (1/N, \Hist[1].\mbox{freq}) + \varepsilon$ \Comment{allowed level}
  \State $\hostload\gets (1 - \sum_i \Hist[i].\mbox{freq})/H$ \Comment{average host load}
\ForAll{keys $k$ with frequency $f$ in $\Hist$}\Comment{heavy keys}
\State $p\gets KI(k)$\par \Comment{try to place $k$ into the same partition as before}
\If{the load of partition $p$ is less than $\maxload-f$}
\State keep $k$ in $p$; increase load of $p$ by $f$; \textbf{continue}
\EndIf
\State $p \gets \Hash(k)$ \Comment{Try the hash location}\par \Comment{to reduce potential migration later}
\If{the load of partition $p$ is less than $\maxload-f$}
\State put $k$ in $p$; increase load of $p$ by $f$; \textbf{continue}
\EndIf
\State Put $k$ explicitly to lowest current load  partition
\EndFor
\ForAll{partitions $p$}
\State 
   Compute the load by adding up the relative
   frequency of the heavy keys in $p$
\State  Add $\hostload$ times
   the number of hosts mapped to $p$ by $KI$ to the load
\EndFor
\ForAll{partitions $p$ with load more than $\maxload$}
\State 
   	Move hosts from $p$ to the first partitions with load
   	below $\maxload - \hostload$
\EndFor
\State \Return the new partitioning function
\end{algorithmic}
\end{algorithm}

The default partitioning option in Flink and Spark is the \textbf{Uniform Hash Partitioning (UHP)}, which yields suboptimal performance in case of data skew. The naive solution of routing keys to partitions explicitly requires a large in-memory routing table and involves computationally heavy bin packing.

Our method, \textbf{Key Isolator Partitioner (KIP)} is a heuristic combination of an explicit hashing for the heaviest keys and a weighted hash partitioner for filling up the partitions to roughly the same load. 
For a repartitioning decision, we also try to make minimal modifications to the previous partitioner to reduce migration costs. This is done by reusing the previous function and rerouting only those heavy keys that would cause imbalance.

In our method, first, we need a distributed top-k histogram computation. While several sampling and heavy hitter identification methods exist in the literature~\cite{gedik2014partitioning}, when experimenting with these methods, we observed either high memory footprint or low performance in improving partitioning balance. For this reason, we implemented a counter-based heuristic algorithm that we describe in our extended paper.

We describe our method to update an existing KIP and prepare for a potential repartitioning in Algorithm~\ref{algorithm:kip}.
Let $N$ be the number of partitions and $\KI$ be the KIP in the previous stage, which maps the keys to one of the $N$ partitions. For keys with no explicit routing, the partition is defined by our weighted hash partitioner $\Hash$, which first maps the keys to one of the $H$ hosts by uniform hashing, and then maps the hosts to partitions.

First, we order the approximate heaviest keys by decreasing frequency in a histogram object \Hist. $\Hist$ is obtained by merging the local histograms that the workers compute during sampling. We only gather the top $B = \lambda N$ keys where $\lambda$ is a global parameter. The $i$-th element of $\Hist$ consists of a key and its relative frequency estimate, $\Hist[i].\mbox{key}$ and $\Hist[i].\mbox{freq}$. Frequency is measured as the fraction of all input, that is, the key frequencies, including also those not in $\Hist$, add up to 1. The ideal maximal load of the partitions with slack $\varepsilon$ is \hbox{$maxload$ = $\max(1/N, \Hist[1].\mbox{freq}) + \varepsilon$.}

In Algorithm~\ref{algorithm:kip}, we first try to keep a heavy key $j$ in its current partition to minimize migration costs. Next, we try $\Hash (j)$, the default location for non-heavy keys, which will be the future location of $j$ in case it becomes non-heavy.  If both of these partitions are full, i.e., if adding key $j$ to either of these partitions increases their current load above $\maxload$, we resort to assigning $j$ to the lowest current load partition.

Next, we deal with all other keys. These keys are handled by the weighted hash partitioner $\Hash$, which first maps the keys to one of the $H\gg N$ hosts and then maps the hosts to partitions. Given no histogram information, we assume that the hosts form a balanced partition of the low frequency keys. The average load of a host, $hostload$, is computed by adding up the frequency estimates of the heavy keys and dividing the remaining frequency by the number of hosts $H$. Formally, $hostload = (1 - \sum_i \Hist[i].\mbox{freq})/H$.  Finally, the hosts are rerouted as necessary by greedy bin packing.

\section{Evaluation}
\label{sect:eval}

First, we analyze our hashing technique, and then we perform speedup experiments of DR in Spark and Flink streaming.
We ran component tests on Intel i7 processors with sample data preloaded into the memory and system tests on on-prem cloud environments. 
We use two datasets, \textbf{LFM} of 4M tags of LastFM music listening records, and \textbf{ZIPF} of 4M element parametrized Zipfian datasets of 100K distinct items, with an exponent between 1--3.

\subsubsection{Evaluation of hashing techniques}

We measured and compared the running time, load balance, and state migration cost of KIP to UHP, to our implementation of partitioning methods \emph{Readj}, \emph{Redist}, and \emph{Scan} from~\cite{gedik2014partitioning}, and to partitioning strategy \emph{Mixed} from~\cite{fang2016parallel}.
We set $\lambda = 2$ for KIP. We run Readj, Redist, and Scan with linear resource functions, balance constraints $\theta_s =\theta_c =\theta_n = 0.2$ and utility function $U = \rho +\gamma$, and Mixed with the same histogram size bound ($A_{max}$) as for KIP and with load balance upper bound $\theta_{\max}$ obtained through an extra optimization loop.  As data, we use LFM and ZIPF of exponent~1. 

\begin{figure}[t]
  \centering
  \includegraphics[width=0.49\columnwidth]{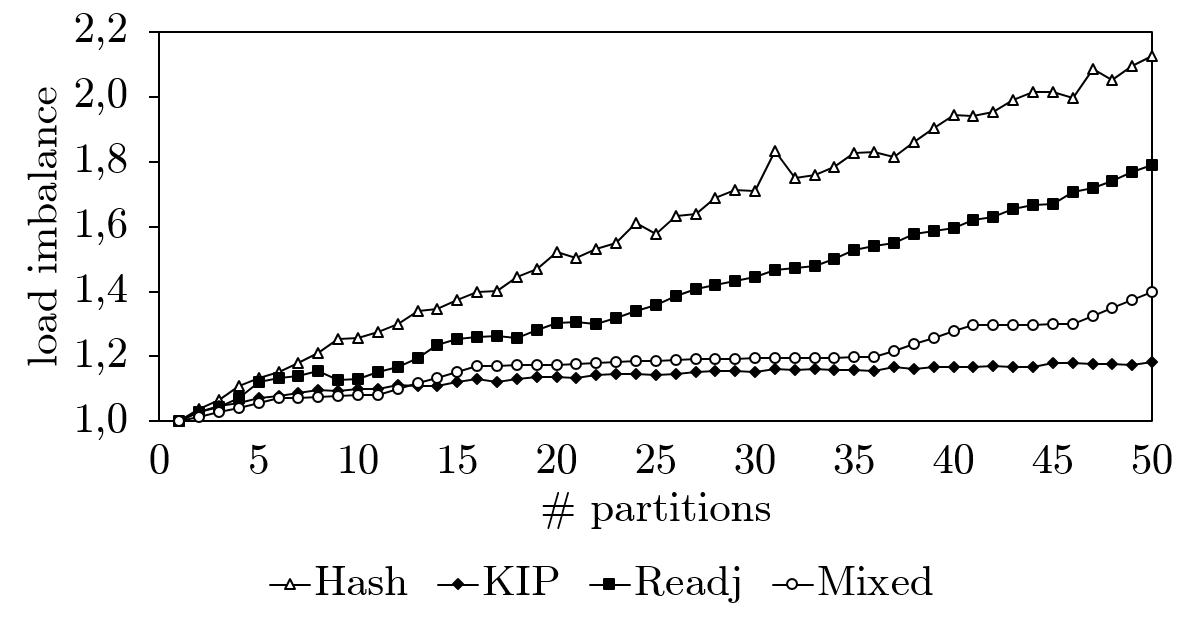}
  \includegraphics[width=0.49\columnwidth]{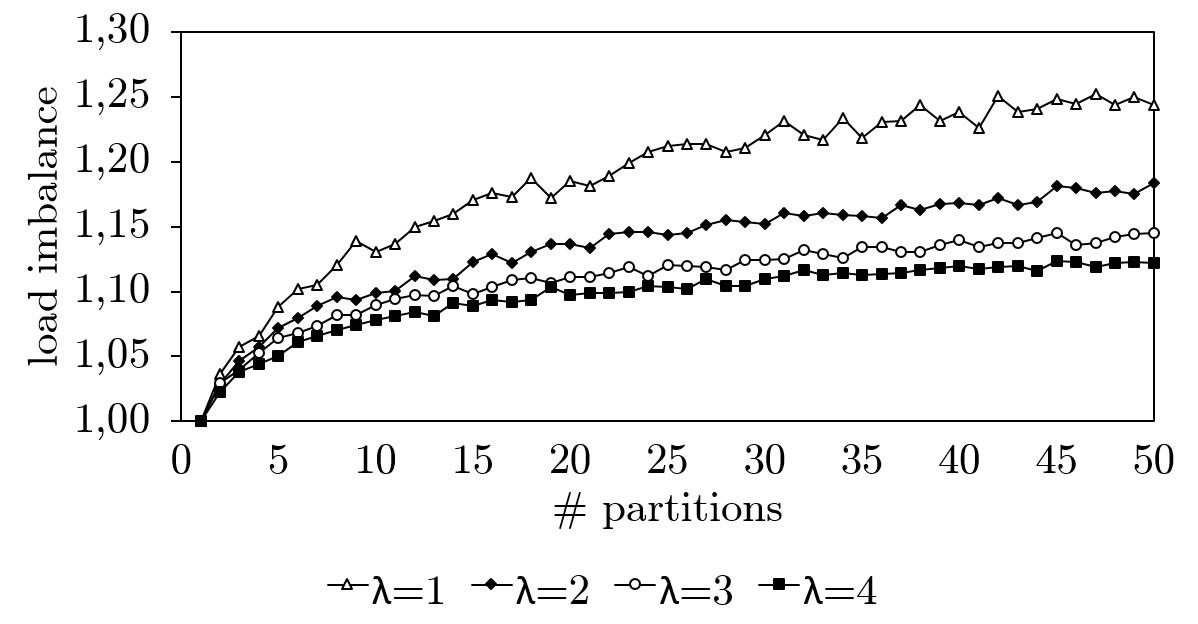}
  \caption{Effect of parallelism on load imbalance over ZIPF of exponent 1 with different partitioning methods (left), and with KIP with varying global histogram scale factor $\lambda$ (right).}
  \label{fig:zipf_imbalance}
\end{figure}

In Figure~\ref{fig:zipf_imbalance}, we investigate load imbalance (the fraction of the maximal load and the average load) over ZIPF as an average of 100 independent experiments. We compare different partitioning methods, as well as KIP with varying global histogram scale factor $\lambda\in\{1, 2, 3, 4\}$. The load imbalance of Hash and Readj grows linearly with the number of partitions; Redist and Scan performs very similarly to Readj. Imbalance grows much slower for Mixed and stays below 1.2 for KIP. Mixed requires a user-supplied load balance upper bound $\theta_{max}$ that we optimized in advance.
KIP reaches better load balance for higher values of $\lambda$. The more heavy keys handled by explicit hashing, the more control KIP has over load balance.

\begin{figure}[t]
   \centering
   \includegraphics[width=.49\columnwidth]{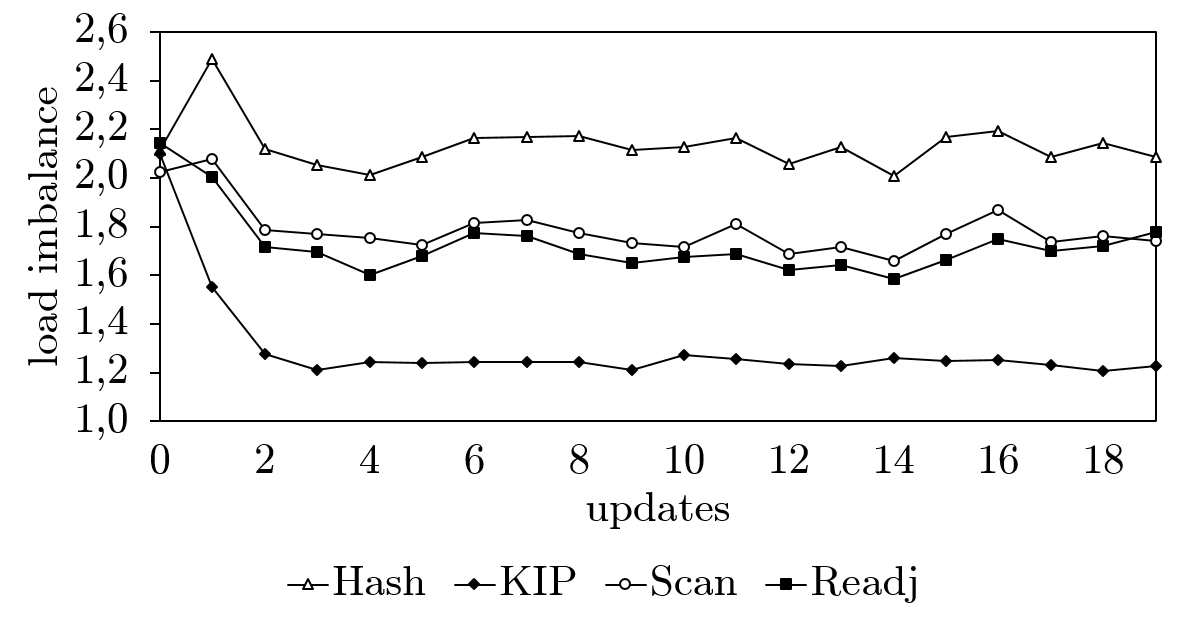}
   \includegraphics[width=.49\columnwidth]{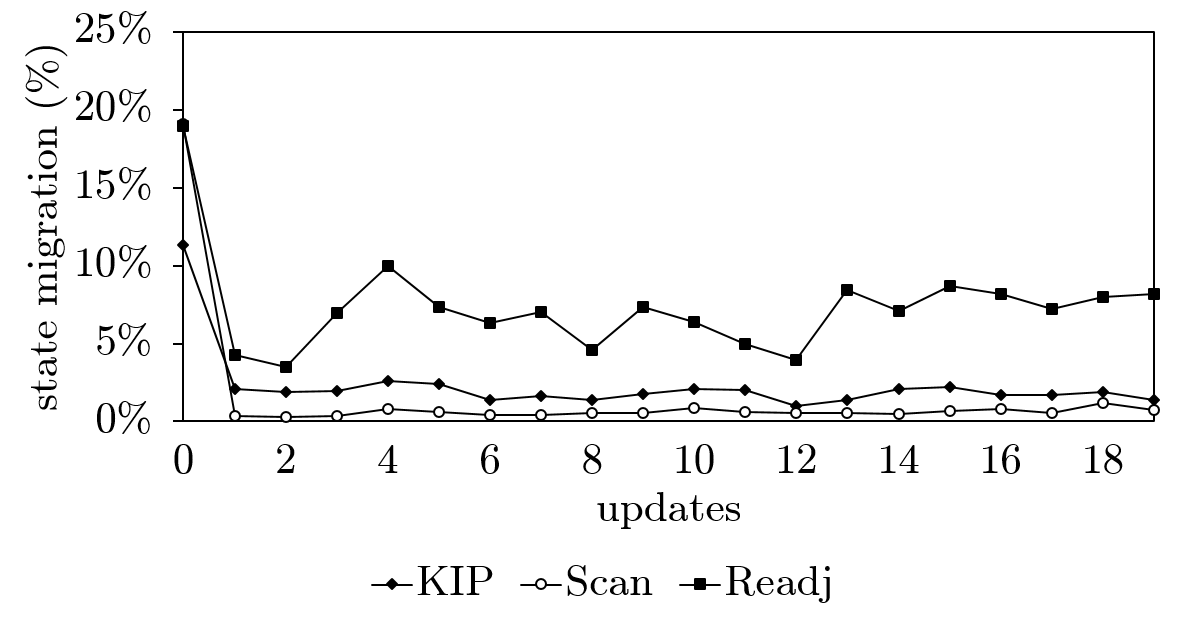}
   \caption{Load imbalance (left) and relative state migration (right) over a stream of LFM consisting of 20 batches, each of size 100K. Update 0 marks the first replacement of UHP.}
   \label{fig:lastfm_updates}
\end{figure}

In the last two experiments in Figure~\ref{fig:lastfm_updates}, we compare load imbalance and relative state migration of KIP to Scan and Readj over LFM. We also plot the load imbalance of Hash for comparison. We omit Redist as it yields very similar results to Scan.
We split LFM into 20 batches of size 100K, and used 20 partitions. States were assumed to be linear in the size of the corresponding key-groups and were kept in a sliding state window of size 5. We forced a partitioner update on each batch. We averaged measurements over 10 iterations, replacing keys with randomly generated strings in each round.
All partitioning methods started with a load imbalance of around 2.0 and a relatively heavy migration caused by switching from Hash to one of the dynamic partitioners.

We conclude that KIP improves load imbalance by 41\%, 29\%, and 26\% compared to Hash, Scan, and Readj, respectively, and handles fluctuations in key distribution much better than the other methods. In terms of state migration, KIP outperforms Readj by a factor of 4 and yields a practically acceptable migration cost, while maintaining consistently lower load imbalance. Scan, which explicitly optimizes migration, performs even better at the cost of load balance.

Our measurements in our extended paper show benefits of KIP in other areas, for example, we confirm that the cost of KIP update is significantly less than that of the other partitioning methods. The detailed parameters of our KIP methods are found in our source code.

\subsubsection{Evaluation of Apache Spark DR}

We experimented with Spark~2.4.0 over a 4-node cluster running Hadoop, with each node being equipped with 10 cores. We generated Zipfian distributions using exponents between 1 and 2. We set the number of keys to be 1M and used the MurmurHash3 algorithm to generate word tokens, including a payload of a timestamp. In a Spark Streaming program, we group events by tokens, then sort them by their time\-stamp, and feed them to an NLP model, a common operation that requires key-grouping.

\begin{figure}[t]
   \centering
   \includegraphics[width=0.48\columnwidth,trim=0 0 0 0, clip]{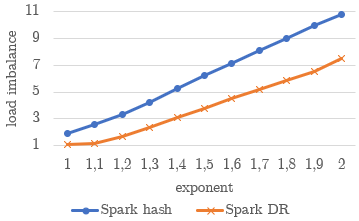}
   \includegraphics[trim={0 0 0 0.1cm},clip,width=0.48\columnwidth]{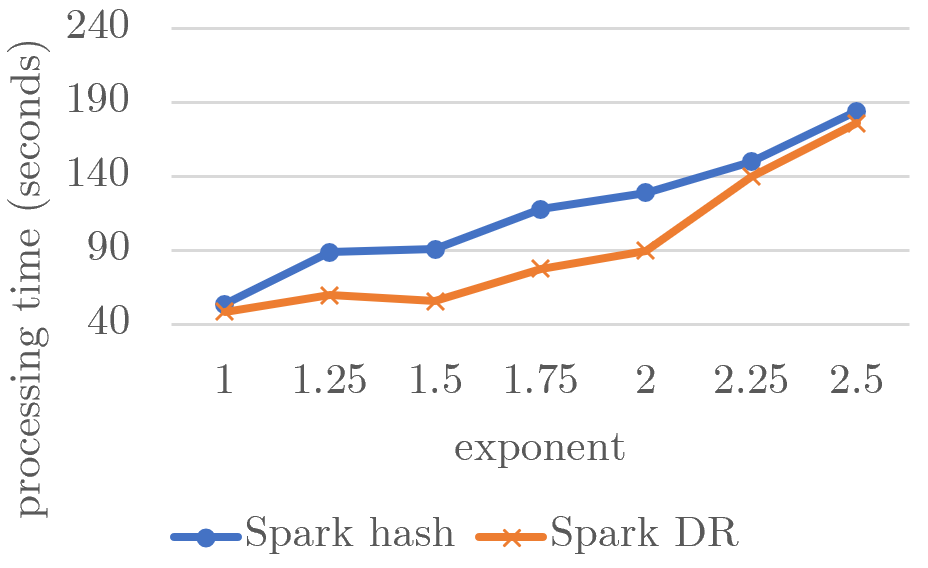}
   \caption{Load imbalance (left) and total Spark processing time (right) for 10M ZIPF records.}
   \label{fig:spark:balance}
\end{figure}

In Figure~\ref{fig:spark:balance}, we show how load balance can be improved with DR, using 35 partitions and an incoming data rate of 50,000 (0.4 GB) per partition.
We also show the time required to process 10M ZIPF records, including the Mapper and Reducer phases as well as any scheduling overhead that comes with over-partitioning.
We observe that DR is beneficial for the moderate values of the Zipf exponent. For an exponent near 1, DR is not required since the key distribution is not significantly skewed, and not even the partitions of the heaviest keys become stragglers. On the other end, for very large exponents, the heaviest key dominates the processing time and other keys randomly hashed to the partition of the heaviest key make little difference in the relative running time of the straggler partition.

\begin{figure}[t]
   \centering
   \includegraphics[width=0.48\columnwidth,trim=0 0 0 0, clip]{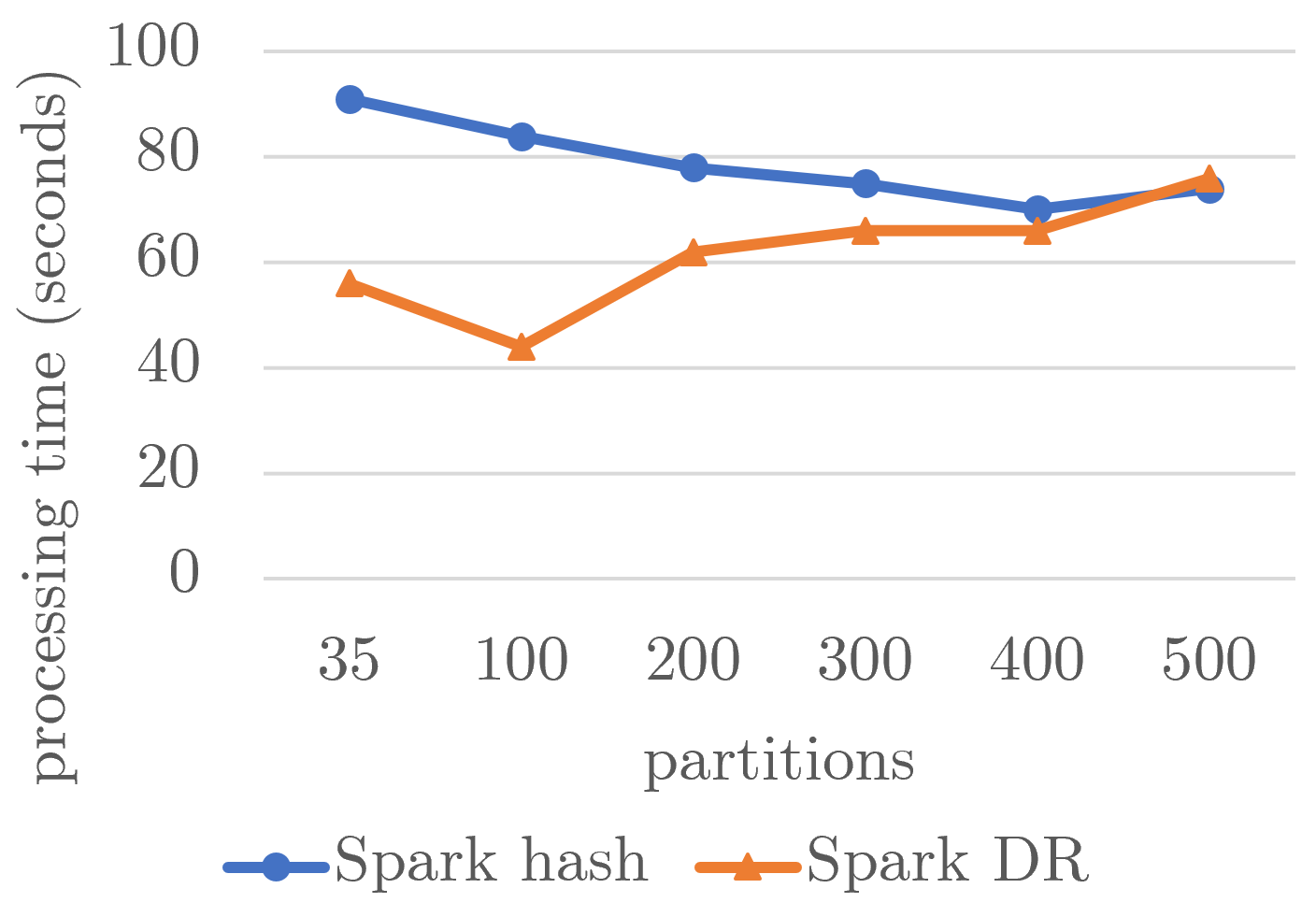}
   \includegraphics[width=0.48\columnwidth]{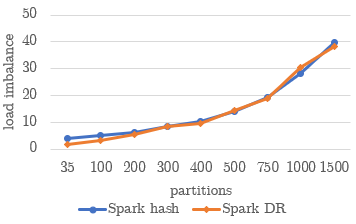}
   \caption{Processing time (left) and load imbalance (right) of Spark with and without DR, over ZIPF data of exponent 1.5, as the function of the number of partitions.}
   \label{fig:spark:partitions}
\end{figure}

Figure~\ref{fig:spark:partitions} shows how DR compares to over-partitioning with ZIPF data of exponent 1.5. The optimal number of partitions for a resource configuration differs in the case of Spark with and without DR. Over-partitioning is beneficial in both cases; DR performs best when the number of partitions is equal to 2--3 times the number of available compute slots. For DR, a higher number of partitions incurs more overhead, while without DR, processing time keeps improving. Nevertheless, we cannot reach the speedup of DR by over-partitioning.

\begin{figure}[t]
   \centering
   \includegraphics[width=0.48\columnwidth, trim=0 0 0 35, clip=true]{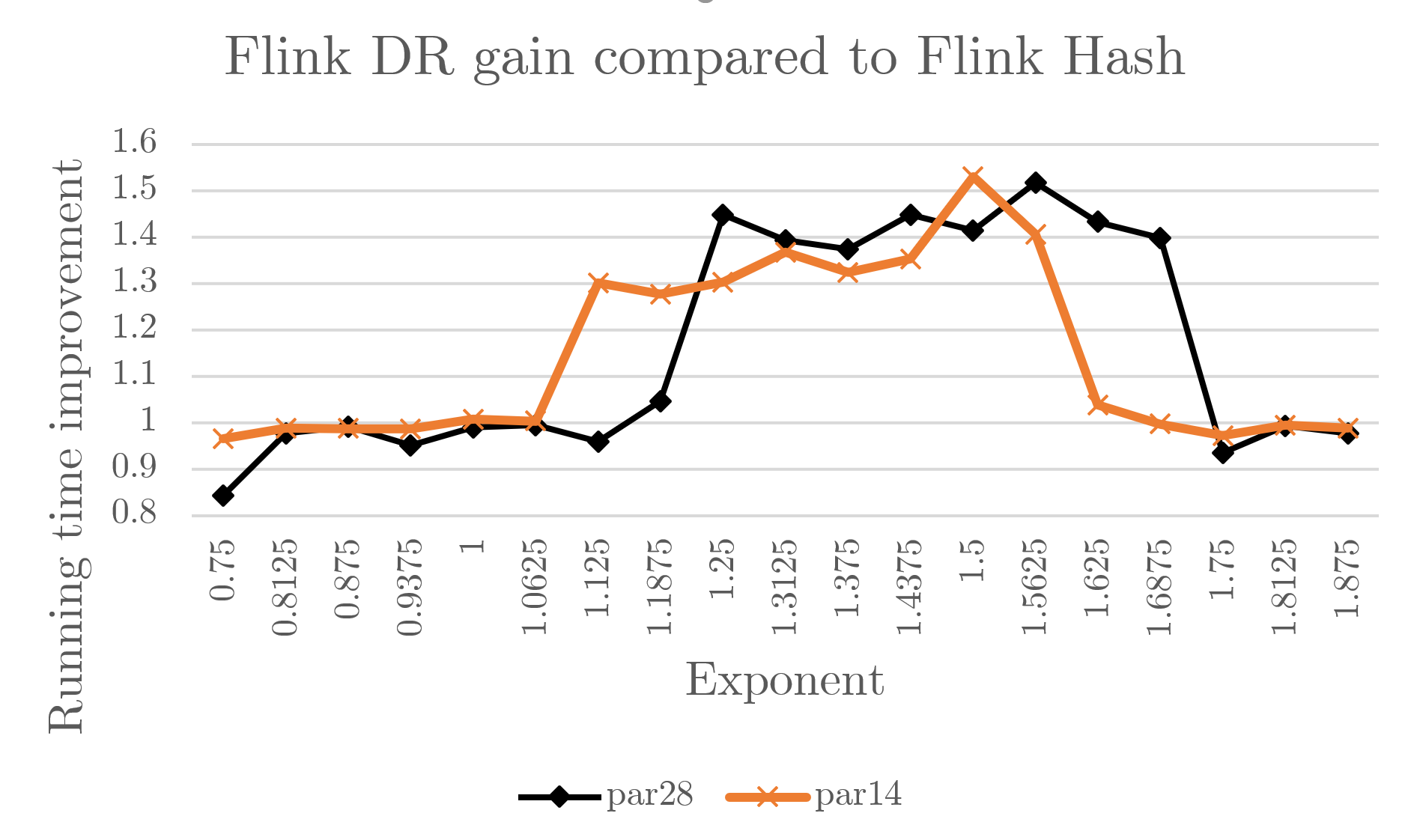}
   \includegraphics[width=0.48\columnwidth]{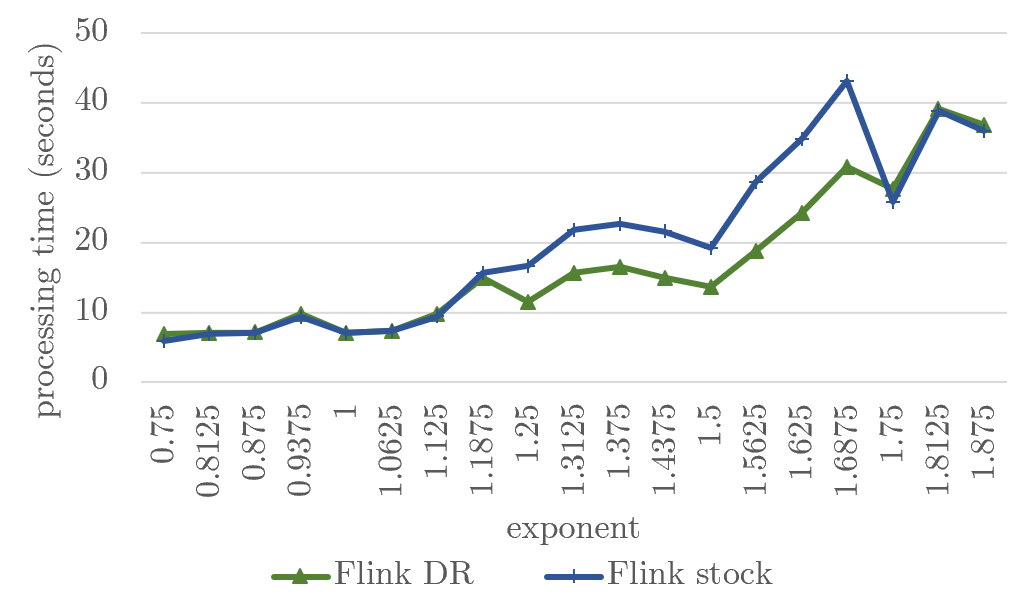}
   \caption{(Left) Relative increase in Flink throughput achieved by DR, with parallelism 14 and 28. (Right) Improvement in running time achieved by Flink with DR compared to Flink without DR, with parallelism 28.}
   \label{fig:eval.flink.running.time.par14}
\end{figure}

\subsubsection{Evaluation of Apache Flink DR}

Measurements of our Flink integration were conducted on a 15-node Dockerized cloud environment with recommended and default configuration, including network buffers. We used Flink~1.3 with 14 \textit{TaskManagers}, each with 4 CPUs and 16 GB RAM, that is, all available resources used.

We measured the effectiveness and the overhead of DR on Zipfian distributions with 1M keys. We used a reducer that simply stores a count for each key as task state.  We experimented with an under-utilized cluster of 14 sources and reducers and a fully utilized cluster of 28 sources and reducers. Each source generated 57,500 (0.5 GB) records per second. 
Note that overpartitioning in Flink is not beneficial. Flink deploys long-running tasks that cannot be scheduled one after another. Hence they compete for resources, which results in performance degradation.
For this reason, we omit the Flink counterpart of the measurements.

Figure~\ref{fig:eval.flink.running.time.par14} shows the relative increase in throughput over Zipfian distributions of different exponents with and without DR. Each measurement represents a throughput measured in the first 10 minutes of the Flink job. Figure~\ref{fig:eval.flink.running.time.par14}
shows the processing time for 10M records with 28 executors. Similar to Spark in Figure~\ref{fig:spark:balance}, we observe improvement for the moderate exponents.

\begin{figure}[t]
   \centering
   \includegraphics[width=0.49\columnwidth,trim=0 0 0 0, clip]{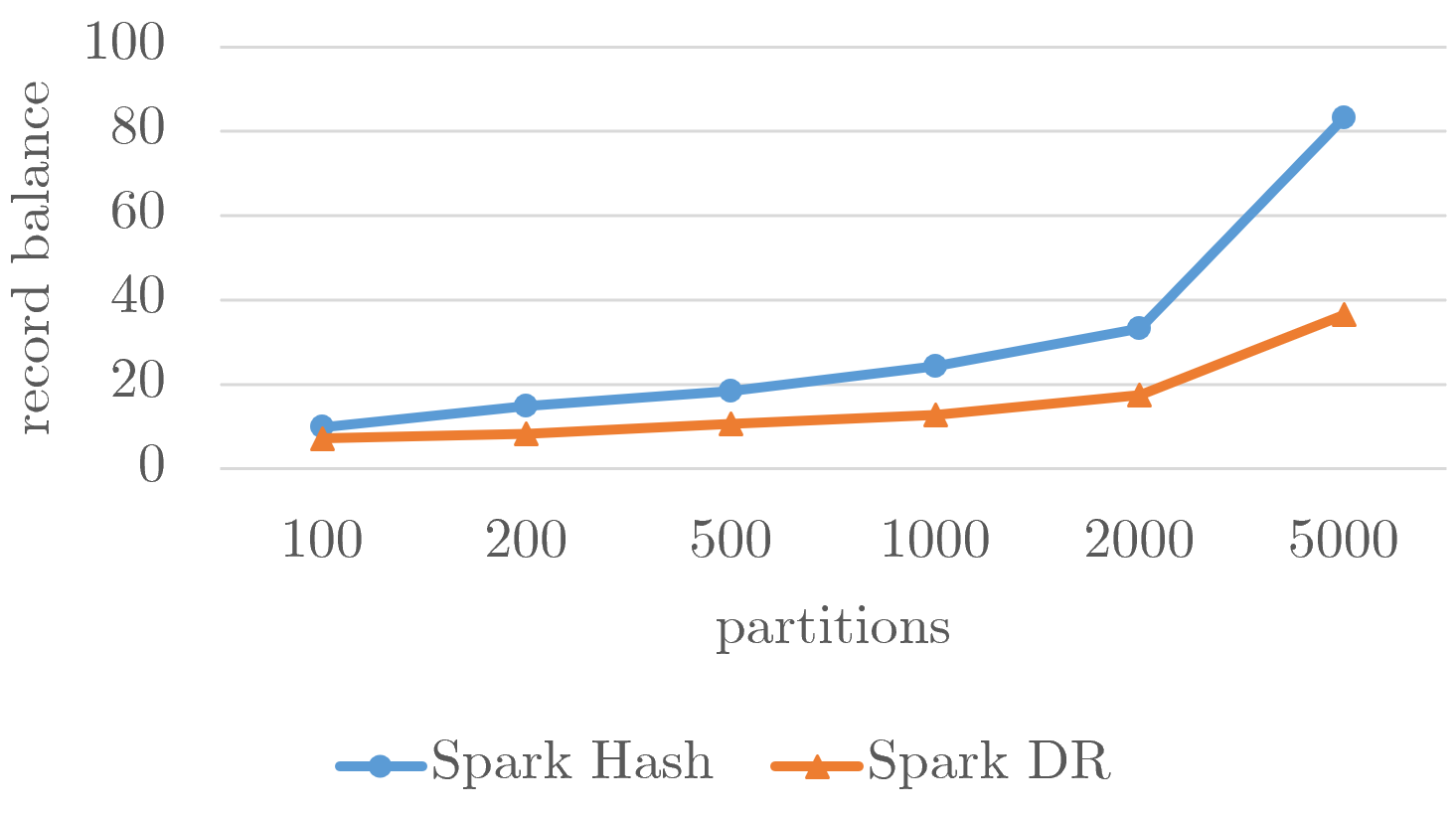}
   \includegraphics[width=0.49\columnwidth]{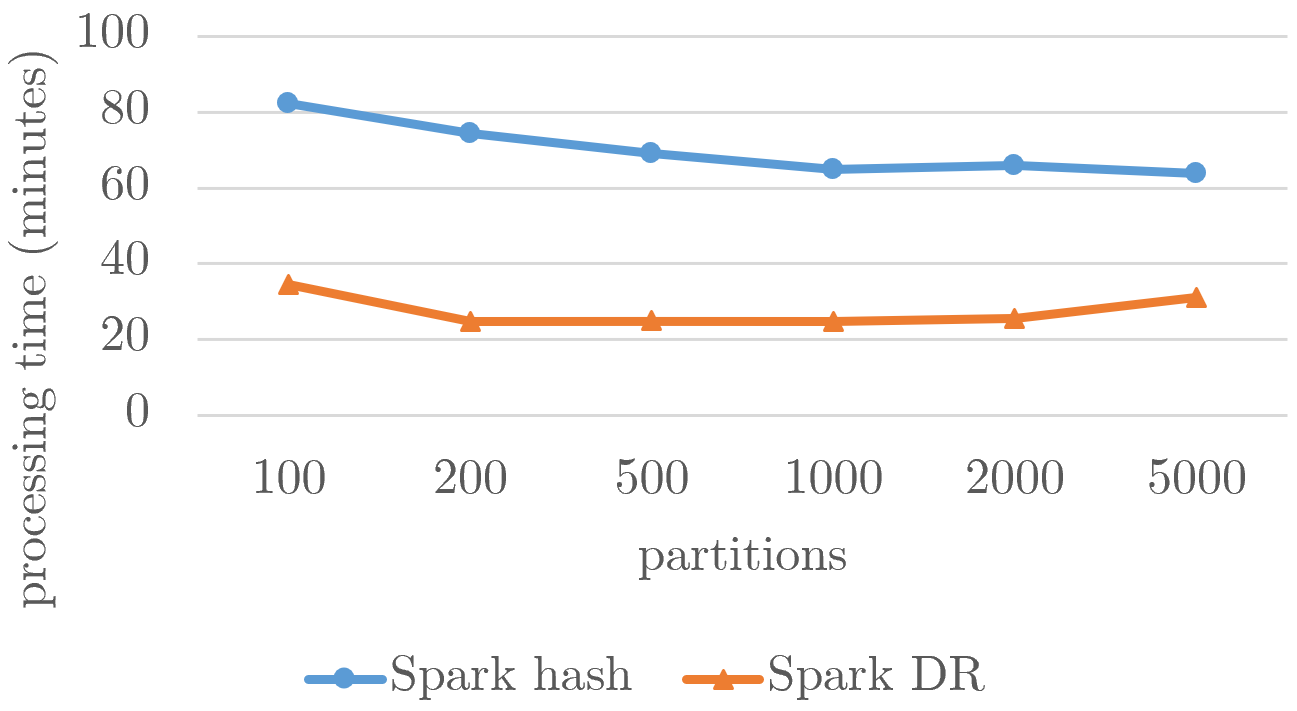}
   \caption{Record balance (left) and processing time (right) of Spark with and without DR, over web crawl data in the 7th round of crawl.}
   \label{fig:spark:crawl}
\end{figure}

\begin{figure}[t]
   \centering
   \includegraphics[width=0.49\columnwidth,trim=0 0 0 0, clip]{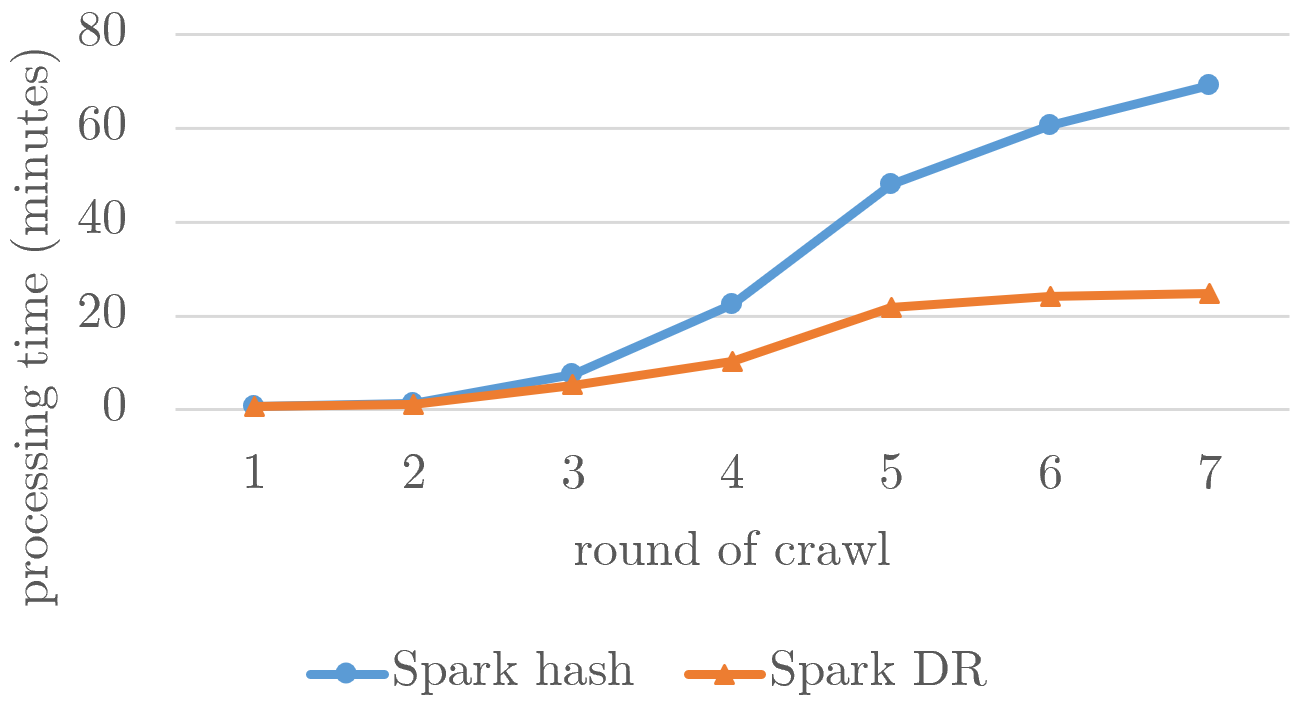}
   \includegraphics[width=0.49\columnwidth]{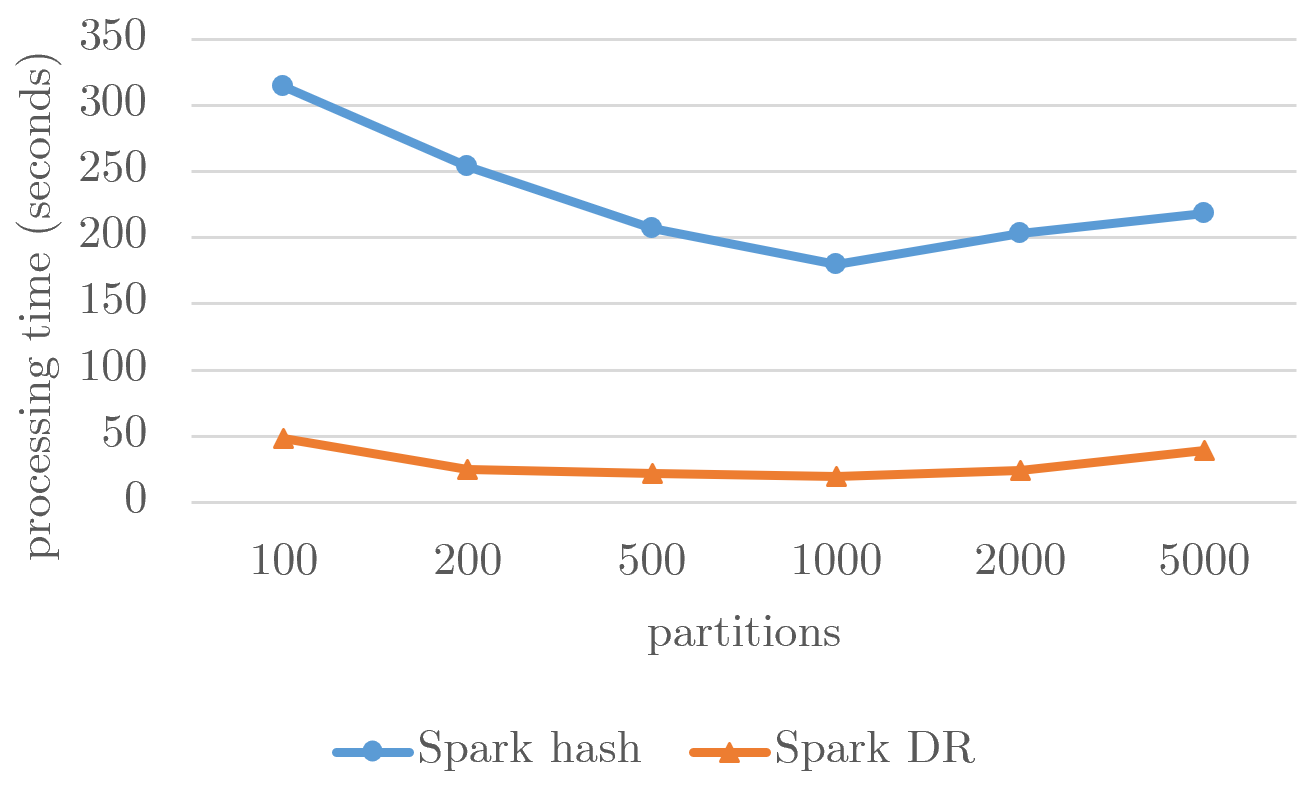}
   
   \caption{(Left) Speedup of Spark DR of consecutive crawl rounds compared to Spark hash. (Right) Processing time of Spark with and without DR, over the NER streaming application.}
   \label{fig:spark:crawl_rounds}
\end{figure}

\section{Web crawl load balancing}
\label{sect:webcrawl}

In this section, we describe a real-world use case of Dynamic Repartitioning. We show how DR can improve fetch list partitioning in a web crawl. As we show, in several cases, web crawl data has to be processed and partitioned by web hosts. The amount of information from hosts has a heavily skewed distribution, which is not necessarily known before starting the crawl.

Next, we show two applications: first, we improve on the hash partitioned web page fetching and processing tasks, and then we perform host-level content analysis by using named entity recognition.

\subsubsection{Balancing in distributed crawl ordering}

We perform web fetching and link extraction partitioned by domains. The reason for domain partitioning stems from the crawler politeness requirements~\cite{olston2010web}. After fetching the elements of a web page, a typically complex step is to process the content to extract text and links. Our crawler implementation maintains a static pool of web browser drivers to fully load dynamic pages of news articles that require JavaScript. This involves heavy processing with an a priori unpredictable heavy-tailed resource distribution depending on the content management technology~\cite{choudhary2012crawling}. 

In our experiment, we injected a list of 64 news sites into the crawler, and to facilitate faster experiments, allowed depth one to crawl domains referenced from the initial domain pool, but no further. Results are similar with increased depth. In each crawl round, we partitioned the fetch lists by host and assigned a web browser instance to load websites dynamically. Then, we used wrapper induction techniques to extract articles.

We measured performance with an 8-node Kubernetes cluster running 8 Spark executors, each with 8 cores and 50 GB of memory. Figure~\ref{fig:spark:crawl} shows how Spark DR improves the balance of fetch lists across all partitions. The running time of the final 7th crawl round (processing of 230 GB of data) is significantly reduced compared to partitioning with Spark's uniform hashing. Figure~\ref{fig:spark:crawl_rounds} shows how DR improves crawl speed in consecutive crawl rounds. On the 7th round, the time required to crawl and parse articles has been reduced from 69.1 to 24.9 minutes.

\subsubsection{Entity recognition in web domains}

In the second use case, we partition the result of a web crawl by hosts for web domain classification. This typical task reaches better accuracy than single page-level classification~\cite{da2006site,castillo2007know}, since pages from the same host are typically strongly inter-related, sometimes even dependent on each other.
By partitioning, we are also capable of domain adaptation by processing and analyzing a large interrelated set of documents in one executor.

In our experiment, we feed the web crawler output into a Spark Streaming application. Then a NER model~\cite{whitelaw2008web} is used to calculate frequent mentions of the recognized entities in 60-minute time windows. Here, we partition by host, since NER tools on the web require domain adaptation~\cite{zhu2005espotter}. Calculating frequent mentions requires sorting of entities within the time window and a mutable update of state per domain key. NLP tools such as named entity recognition are sensitive to the length of text, therefore certain domains require increased processing time.

Experiments were conducted on a Kubernetes cluster of 6 nodes, running 6 Spark executors, each with 6 cores and 15 GB of memory. Figure~\ref{fig:spark:crawl_rounds} shows the time required to process a reference of 40,000 records (0.9 GB). In our experiments, DR was capable of speeding up the completion of the NER task by a factor of 6 for all partition configurations. The reason for strong gains compared to the simple counting tasks is that the map-reduce tasks require heavy processing with large states. Such tasks, in general, can greatly benefit from DR.

\subsubsection*{Conclusions}

We described our design and architecture of a lightweight dynamic repartitioning, which is universally adaptable for distributed batch and stream data processing systems.  The novelty of our system lies in the possibility to handle long-running stateful streaming tasks as well.
By interacting with the underlying system components, our system approximately computes key histograms, heuristically determines adaptive partitioning strategies on the fly, and injects the new partitioner either between micro-batches or at streaming checkpoints.
We measured our solution over Spark and Flink. We reached a 1.5 to 6-times speedup with synthetic power-law key distributions and with real Spark-based web crawling tasks.
Our source codes and extended paper are available at \url{https://github.com/zzvara/dynamic-repartitioning-paper}.

\bibliographystyle{splncs04}
\bibliography{paper}

\begin{thebibliography}{10}
\providecommand{\url}[1]{\texttt{#1}}
\providecommand{\urlprefix}{URL }
\providecommand{\doi}[1]{https://doi.org/#1}

\bibitem{abadi2005design}
Abadi, D.J., Ahmad, Y., Balazinska, M., Cetintemel, U., Cherniack, M., Hwang,
  J.H., Lindner, W., Maskey, A., Rasin, A., Ryvkina, E., et~al.: The design of
  the borealis stream processing engine. In: CIDR. vol.~5, pp. 277--289 (2005)

\bibitem{bifet2011data}
Bifet, A., Kirkby, R., Pfahringer, B.: Data stream mining: a practical
  approach. Tech. rep., University of Waikato (2011)

\bibitem{carbone2015lightweight}
Carbone, P., Fóra, G., Ewen, S., Haridi, S., Tzoumas, K.: Lightweight
  asynchronous snapshots for distributed dataflows (2015)

\bibitem{castillo2007know}
Castillo, C., Donato, D., Gionis, A., Murdock, V., Silvestri, F.: Know your
  neighbors: Web spam detection using the web topology. In: Proc. 30th SIGIR.
  pp. 423--430. ACM (2007)

\bibitem{choudhary2012crawling}
Choudhary, S., Dincturk, M.E., Mirtaheri, S.M., Moosavi, A., Von~Bochmann, G.,
  Jourdan, G.V., Onut, I.V.: Crawling rich internet applications: the state of
  the art. In: Proc. of the 2012 Conference of the Center for Advanced Studies
  on Collaborative Research. pp. 146--160. IBM Corp. (2012)

\bibitem{da2006site}
da~Costa~Carvalho, A.L., Chirita, P.A., De~Moura, E.S., Calado, P., Nejdl, W.:
  Site level noise removal for search engines. In: Proc. 15th WWW. pp. 73--82.
  ACM (2006)

\bibitem{dewitt1992practical}
DeWitt, D.J., Naughton, J.F., Schneider, D.A., Seshadri, S.: Practical skew
  handling in parallel joins. UW-Madison. Computer Sciences Department (1992)

\bibitem{ding2015optimal}
Ding, J., Fu, T.Z., Ma, R.T., Winslett, M., Yang, Y., Zhang, Z., Chao, H.:
  Optimal operator state migration for elastic data stream processing. arXiv
  preprint arXiv:1501.03619  (2015)

\bibitem{fang2016parallel}
Fang, J., Zhang, R., Fu, T.Z., Zhang, Z., Zhou, A., Zhu, J.: Parallel stream
  processing against workload skewness and variance. arXiv:1610.05121  (2016)

\bibitem{gama2013evaluating}
Gama, J., Sebasti{\~a}o, R., Rodrigues, P.P.: On evaluating stream learning
  algorithms. Machine learning  \textbf{90}(3),  317--346 (2013)

\bibitem{gates2013apache}
Gates, A., Dai, J., Nair, T.: Apache pig's optimizer. IEEE Data Engineering
  Bulletin  \textbf{36}(1),  34--45 (2013)

\bibitem{gedik2014partitioning}
Gedik, B.: Partitioning functions for stateful data parallelism in stream
  processing. The VLDB Journal  \textbf{23}(4),  517--539 (2014)

\bibitem{gufler2012partition}
Gufler, B., Augsten, N., Reiser, A., Kemper, A.: The partition cost model for
  load balancing in mapreduce. In: Cloud Computing and Services Science, pp.
  371--387. Springer (2012)

\bibitem{hidalgo2017self}
Hidalgo, N., Wladdimiro, D., Rosas, E.: Self-adaptive processing graph with
  operator fission for elastic stream processing. JSS  \textbf{127},  205--216
  (2017)

\bibitem{hirzel2014catalog}
Hirzel, M., Soul{\'e}, R., Schneider, S., Gedik, B., Grimm, R.: A catalog of
  stream processing optimizations. ACM Computing Surveys (CSUR)
  \textbf{46}(4), ~46 (2014)

\bibitem{hulten2001mining}
Hulten, G., Spencer, L., Domingos, P.: Mining time-changing data streams. In:
  Proc. 7th SIGKDD. pp. 97--106. ACM (2001)

\bibitem{katsipoulakis2017holistic}
Katsipoulakis, N.R., Labrinidis, A., Chrysanthis, P.K.: A holistic view of
  stream partitioning costs. Proc. VLDB  \textbf{10}(11),  1286--1297 (2017)

\bibitem{kwon2012skewtune}
Kwon, Y., Balazinska, M., Howe, B., Rolia, J.: Skewtune: mitigating skew in
  mapreduce applications. In: Proc. SIGMOD. pp. 25--36. ACM (2012)

\bibitem{kwon2013managing}
Kwon, Y., Ren, K., Balazinska, M., Howe, B., Rolia, J.: Managing skew in
  hadoop. IEEE Data Engineering Bulletin  \textbf{36}(1),  24--33 (2013)

\bibitem{lam2012muppet}
Lam, W., Liu, L., Prasad, S., Rajaraman, A., Vacheri, Z., Doan, A.: Muppet:
  Mapreduce-style processing of fast data. Proc. VLDB  \textbf{5}(12),
  1814--1825 (2012)

\bibitem{manku2002approximate}
Manku, G.S., Motwani, R.: Approximate frequency counts over data streams. In:
  Proc. of the 28th international conference on Very Large Data Bases. pp.
  346--357. VLDB Endowment (2002)

\bibitem{metwally2005efficient}
Metwally, A., Agrawal, D., El~Abbadi, A.: Efficient computation of frequent and
  top-k elements in data streams. In: International Conference on Database
  Theory. pp. 398--412. Springer (2005)

\bibitem{metwally2012v}
Metwally, A., Faloutsos, C.: V-smart-join: A scalable mapreduce framework for
  all-pair similarity joins of multisets and vectors. Proc. VLDB
  \textbf{5}(8),  704--715 (2012)

\bibitem{nasir2015partial}
Nasir, M.A.U., Morales, G.D.F., Garcia-Soriano, D., Kourtellis, N., Serafini,
  M.: Partial key grouping: Load-balanced partitioning of distributed streams.
  arXiv preprint arXiv:1510.07623  (2015)

\bibitem{nasir2016two}
Nasir, M.A.U., Morales, G.D.F., Kourtellis, N., Serafini, M.: When two choices
  are not enough: Balancing at scale in distributed stream processing. In: 2016
  IEEE 32nd International Conference on Data Engineering. pp. 589--600. IEEE
  (2016)

\bibitem{neumeyer2010s4}
Neumeyer, L., Robbins, B., Nair, A., Kesari, A.: S4: Distributed stream
  computing platform. In: 2010 IEEE ICDMW. pp. 170--177 (2010)

\bibitem{olston2010web}
Olston, C., Najork, M., et~al.: Web crawling. Foundations and
  Trends{\textregistered} in Information Retrieval  \textbf{4}(3),  175--246
  (2010)

\bibitem{rivetti2015proactive}
Rivetti, N., Anceaume, E., Busnel, Y., Querzoni, L., Sericola, B.: Proactive
  online scheduling for shuffle grouping in distributed stream processing
  systems. Ph.D. thesis, LINA-University of Nantes; Sapienza Universit{\`a} di
  Roma (Italie) (2015)

\bibitem{rivetti2015efficiently}
Rivetti, N., Busnel, Y., Most{\'e}faoui, A.: Efficiently summarizing data
  streams over sliding windows. In: 2015 IEEE NCA. pp. 151--158 (2015)

\bibitem{schneider1989performance}
Schneider, D.A., DeWitt, D.J.: A performance evaluation of four parallel join
  algorithms in a shared-nothing multiprocessor environment. In: Proc. SIGMOD.
  pp. 110--121 (1989)

\bibitem{schneider2015safe}
Schneider, S., Hirzel, M., Gedik, B., Wu, K.L.: Safe data parallelism for
  general streaming. IEEE Transactions on Computers  \textbf{64}(2),  504--517
  (2015)

\bibitem{toshniwal2014storm}
Toshniwal, A., Taneja, S., Shukla, A., Ramasamy, K., Patel, J.M., Kulkarni, S.,
  Jackson, J., Gade, K., Fu, M., Donham, J., et~al.: {Storm @ Twitter}. In:
  Proc. SIGMOD. pp. 147--156 (2014)

\bibitem{vernica2012adaptive}
Vernica, R., Balmin, A., Beyer, K.S., Ercegovac, V.: Adaptive mapreduce using
  situation-aware mappers. In: Proc. of the 15th International Conference on
  Extending Database Technology. pp. 420--431. ACM (2012)

\bibitem{whitelaw2008web}
Whitelaw, C., Kehlenbeck, A., Petrovic, N., Ungar, L.: Web-scale named entity
  recognition. In: Proc. of the 17th ACM conference on Information and
  knowledge management. pp. 123--132. ACM (2008)

\bibitem{xing2005dynamic}
Xing, Y., Zdonik, S., Hwang, J.H.: Dynamic load distribution in the borealis
  stream processor. In: Proc. ICDE. pp. 791--802. IEEE (2005)

\bibitem{zaharia2010spark}
Zaharia, M., Chowdhury, M., Franklin, M.J., Shenker, S., Stoica, I.: Spark:
  Cluster computing with working sets. HotCloud  \textbf{10}(10-10), ~95 (2010)

\bibitem{zhang2015memory}
Zhang, H., Chen, G., Ooi, B.C., Tan, K.L., Zhang, M.: In-memory big data
  management and processing: A survey. IEEE Transactions on Knowledge and Data
  Engineering  \textbf{27}(7),  1920--1948 (2015)

\bibitem{zhu2005espotter}
Zhu, J., Uren, V., Motta, E.: Espotter: Adaptive named entity recognition for
  web browsing. In: Biennial Conference on Professional Knowledge
  Management/Wissensmanagement. pp. 518--529. Springer (2005)

\bibitem{zliobaite2012next}
{\v{Z}}liobaite, I., Bifet, A., Gaber, M., Gabrys, B., Gama, J., Minku, L.,
  Musial, K.: Next challenges for adaptive learning systems. ACM SIGKDD
  Explorations Newsletter  \textbf{14}(1),  48--55 (2012)

\end{thebibliography}

\end{document}